\documentclass[epsf,twocolumn,showpacs,preprintnumbers]{revtex4}
\usepackage{graphics}
\usepackage{graphicx}
\usepackage{dcolumn} 
\usepackage{bm}
\usepackage{color}
\usepackage{epsfig}
\newcommand{\pco}{PbCrO$_3$}
\newcommand{\sco}{SrCrO$_3$}
\newcommand{\cco}{CaCrO$_3$}
\begin{document}
\title{Orbital-ordering driven structural distortion in metallic \sco.} 
\author{K.-W. Lee$^{1,2}$ and W. E. Pickett$^1$} 
\affiliation{$^1$Department of Physics, University of California, Davis, 
  CA 95616, USA \\
 $^2$Department of Display and Semiconductor Physics, Korea University,
  Jochiwon, Chungnam 339-700, Korea}
\date{\today}
\begin{abstract}
In contrast to the previous reports that the divalent perovskite SrCrO$_3$ 
was believed to be cubic structure and nonmagnetic metal,
recent measurements suggest coexistence of majority tetragonally distorted
weak antiferromagnetic phase and minority nonmagnetic cubic phase.
Within the local (spin) density approximation (L(S)DA) our calculations confirm 
that a slightly tetragonally distorted phase indeed is energetically
favored.
Using the correlated band theory method (LDA+ Hubbard $U$) as seems to be justified
by the unusual behavior observed in \sco, above the critical value $U_c$=4 eV
only the distorted phase undergoes an orbital-ordering transition,
resulting in $t_{2g}^2 \rightarrow d_{xy}^1$($d_{xz}d_{yz}$)$^1$ corresponding to
the filling of the $d_{xy}$ orbital but leaving the other two degenerate. 
The Fermi surfaces of the cubic phase are simple with 
nesting features, although the nesting wavevectors do not correlate with known data. 
This is not uncommon in perovskites; the strongly directional $d-d$ bonding often
leads to box-like Fermi surfaces, and either the nesting is not strong
enough, or the matrix elements are not large enough, to promote instabilities. 
Fixed spin moment
calculations indicate the cubic structure is just beyond a ferromagnetic
Stoner instability 
($IN(0) \approx$1.1) in L(S)DA, and that the energy is unusually
weakly dependent on the moment out to 1.5$\mu_B$/Cr (varying only by
11 meV/Cr), reflecting low energy long-wavelength magnetic fluctuations.
We observe that this system shows strong magneto-phonon coupling
(change in Cr local moment is $\sim$7.3 $\mu_B$/\AA)
for breathing phonon modes.
\end{abstract}
\maketitle

\section{Introduction}
Forty years ago, a few divalent chromate perovskites $\cal{A}$CrO$_3$ 
($\cal{A}$=Pb, Sr, Ca),
formally possessing the Cr$^{4+}$ ion, were synthesized 
at high temperature $\sim$1300 K and under high pressure 6--10 GPa
by a few groups.\cite{roth,chamb1,chamb2,good1,chamb3}
In spite of their atypical and controversial properties,
these systems have been little studied, probably due to difficulty of synthesis.
More recently a few groups have begun to revisit the \cco~ and 
\sco~ compounds.\cite{good2,attfield1,attfield2,khomskii1}
Whether these systems are metallic, strongly correlated, and spin ordered
is still controversial.\cite{good2,attfield1,attfield2,khomskii1,khomskii2}

Roth and DeVries reported an ordered moment of 1.9 $\mu_B$
and Curie-Weiss moment of 2.83 $\mu_B$ in the isovalent compound \pco,
consistent with $S=1$ Cr$^{4+}$ ($d^2$).\cite{roth}
Chamberland and Moeller synthesized a single crystal, which was semiconducting
with 0.27 eV activation energy.\cite{chamb1}
There was an anomaly at $T_1$=240 K, and an upturn at $T_2$=160 K,
in susceptibility. The latter was thought to imply a $G$-type antiferromagnetic (AFM)
ordering corresponding to antiparallel spin ordering between all nearest neighbor Cr$^{4+}$
ions. Additionally, at $T_3$=100 K, the logarithmic resistivity shows
a kink, implying another transition.
The samples of both groups had a cubic structure with lattice constant
$a \approx$~4.00~\AA.
Local (spin) density approximation (L(S)DA) calculations 
obtained a magnetic moment of 1.4 $\mu_B$, three-quarters of 
the experimental value, but no band-gap.\cite{jaya}
This difference points to interaction effects beyond those described by L(S)DA.
Chamberland also synthesized a cubic \sco~ with $a$=3.818~\AA,
and concluded it to be a nonmagnetic (NM) metal.\cite{chamb2}

Goodenough {\it et al.} synthesized polycrystalline \cco, which is
orthorhombic ($a$=5.287~\AA, $b$=5.316~\AA, and $c$=7.486~\AA) and
non-conducting (although probably due to polycrystallinity).\cite{good1}
Weiher, Chamberland, and Gillson obtained a metallic single crystal
sample.\cite{chamb3}
The susceptibility measurements showed two anomalies,
a kink at 325 K and an upturn at 90 K. 
At the latter, which is recently identified as the Neel temperature $T_N$,\cite{good2,khomskii1} 
a kink in the resistivity data and decrease in volume by 2 \% 
from the powder diffraction studies were observed.
The Curie-Weiss moment is high spin 3.7 $\mu_B$, recently confirmed by Zhou
{\it et al}.\cite{good2} 
The reason of large difference from spin-only value of 
2.8 $\mu_B$ for $S$=1 system is unresolved, however it was found that the 
susceptibility in \sco~did not follow a Curie-Weiss behavior so no local
moment value could be identified. 

In more recent studies, Zhou {\it et al.} observed a smooth decrease 
in thermal conductivity of \cco~and \sco~ compounds as temperature is lowered, 
in their interpretation characteristic of neither an insulator nor a metal.\cite{good2}
They interpreted this unusual behavior as due to some unusual
Cr--O bonding instability, supported by an increase in compressibility observed
around 4 GPa. 
In contrast to Chamberland's initial suggestion,\cite{chamb2} 
Zhou {\it et al.} concluded that \sco~ is NM and insulating.
Komarek {\it et al.} reported antiferromagnetism with ordering
wavevector $Q_M=(\frac{1}{2},\frac{1}{2},0)$ (in units of $\frac{2\pi}{a}$)
and a saturation magnetic moment of 1.2 $\mu_B$,
using SQUID susceptibility and neutron diffraction.\cite{khomskii1}
At $T_N$, in contrast to the preliminary observations,\cite{chamb3}  
no change in volume was apparent.
They suggested \cco~ is itinerant, but close to being localized, implying
importance of correlation effects.\cite{khomskii1,khomskii2}
Additionally, no evidence of orbital ordering within the $t_{2g}$
shell was observed.\cite{khomskii1} 
 
Attfield and coworkers have concentrated on \sco, using
neutron diffraction and synchrotron powder x-ray diffraction 
studies.\cite{attfield1,attfield2}
Below $T_N \approx$40 K, most of their sample underwent a structure transition
from a NM cubic phase to a AFM tetragonal phase, but the two phases coexist 
even at low $T$. 
In the tetragonal phase with only slightly inequivalent lattice parameters, at $T_N$
there are no visible changes in volume or in averaged Cr-O distance,
and additionally no kink in resistivity. 
The temperature dependent neutron diffraction data imply orbital reoccupation
($d_{xy}d_{xz}d_{yz}$)$^2 \rightarrow$$d_{xy}^1$($d_{xz}d_{yz}$)$^1$ 
in the tetragonal phase at or near $T_N$.
Both phases are metallic, though showing high resistivity due to grain boundary scattering.

In this paper, we will focus on \sco, which has been
little studied theoretically to date.
In Sec. III, in particular, we will address electronic structures 
of both NM cubic phase and AFM distorted phase, 
and show that the inclusion of correlation effects 
leads to an orbital-ordering driven distortion while remaining metallic.
In Sec. IV, we discuss the oxygen breathing vibration that
shows quite strong magneto-phonon coupling.

\section{Structure and Calculation}
Attfield and coworkers suggested a small structure distortion 
involving relative displacement of Sr and O ions, 
leading to $\sqrt{2}a \times \sqrt{2}a\times 2a$ quadrupled supercell
(space group: $Imma$, No. 74).\cite{attfield1}
The corresponding AFM order we consider, which is 
$(\frac{1}{2},\frac{1}{2},\frac{1}{2})$ order in terms of the original perovskite
cell, is a G-type AFM.
Through more precise measurements, 
instead of the quadrupled supercell they more recently concluded 
a tetragonally distorted structure with $c/a \approx$0.992 but 
negligible change in volume (less than 1\%),
which shows antiferromagnetic ordering at wavevector
$Q_M$=($\frac{1}{2},\frac{1}{2}$,0)
(space group: $P4/mmm$, No. 123).\cite{attfield2}
This AFM order is conventionally called C-type AFM.
Although this tetragonal phase is dominant below $T_N$, 
a NM cubic phase also coexists,
consistent with results of our calculations (see below).

We investigated both distorted structures as well as the cubic phase,
using L(S)DA and the LDA+Hubbard $U$ (LDA+U) method.\cite{amf,fll}
For the quadrupled supercell, 
only the planar O position is displaced along the $\langle 001 \rangle$ 
direction, since negligible displacements in Sr and apical O 
were initially suggested.\cite{attfield1}
We used the recent experimental lattice constant $a$=3.811 \AA~
for the cubic structure and for the quadrupled supercell, 
and $a$=3.822 and $c$=3.792 \AA~ for the tetragonal structure.\cite{attfield2}
Our optimized lattice constants in the cubic phase 
within L(S)DA are 3.748 \AA~ for NM and 3.76 \AA~ for AFM, 
reflecting the usual small increase in volume associated with magnetism
and the overbinding that is common in L(S)DA.

We have used the full-potential local orbital code FPLO for our study.
In FPLO-5,\cite{fplo} basis orbitals were chosen as 
Cr $(3s3p)4s4p3d$, Sr $(4s4p)5s5p4d$, and O $2s2p3d$.
(The orbitals in parentheses indicate semicore orbitals.)
The Brillouin zone was sampled with a regular mesh containing 726
irreducible $k$ points, since a fine mesh is required for sampling the
Fermi surface.

\section{Electronic Structure}

\subsection{NM Cubic Phase within LDA}
\begin{figure}[tbp]
\resizebox{8cm}{6cm}{\includegraphics{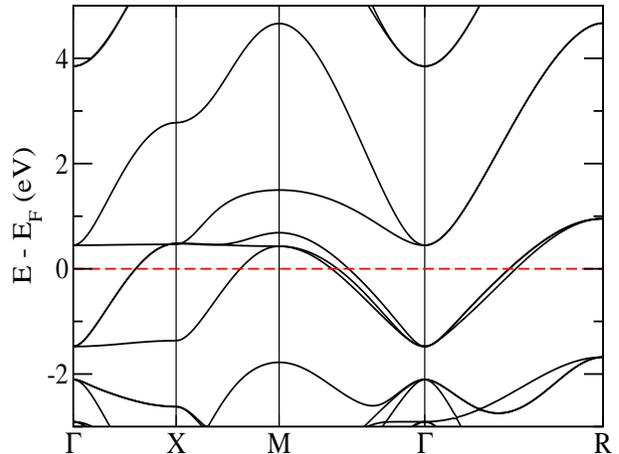}}
\caption{(Color online) Enlarged band structure of NM \sco~
 in the region having mostly Cr $d$ character.
 A flat band along the $\Gamma -X-M$ line lies at 0.5 eV.
 The O $2p$ states lie on the regime of --7.5 to --1.8 eV (not shown here).
 These symmetry points follow a simple cubic notation (see Fig. \ref{FS}).
 The $R$ point is a zone boundary along $\langle111\rangle$ direction.
 The horizontal dashed line indicates the Fermi energy $E_F$.
}
\label{band}
\end{figure}

\begin{figure}[tbp]
\rotatebox{-90}{\resizebox{6.5cm}{7.5cm}{\includegraphics{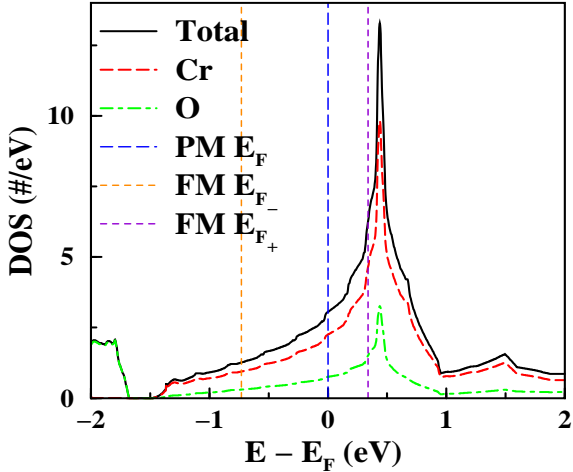}}}
\caption{(Color online) Total and atom-projected densities of states
 in NM \sco.
 A sharp peak lies at 0.5 eV, while around E$_F$ there is only a
 smoothly monotonic behavior.
 The DOS at $E_F$ N(0) is 1.53 states per eV per spin,
 a quarter of which is contributed by O ions.
}
\label{dos}
\end{figure}

\begin{figure}[tbp]
\resizebox{6.5cm}{6cm}{\includegraphics{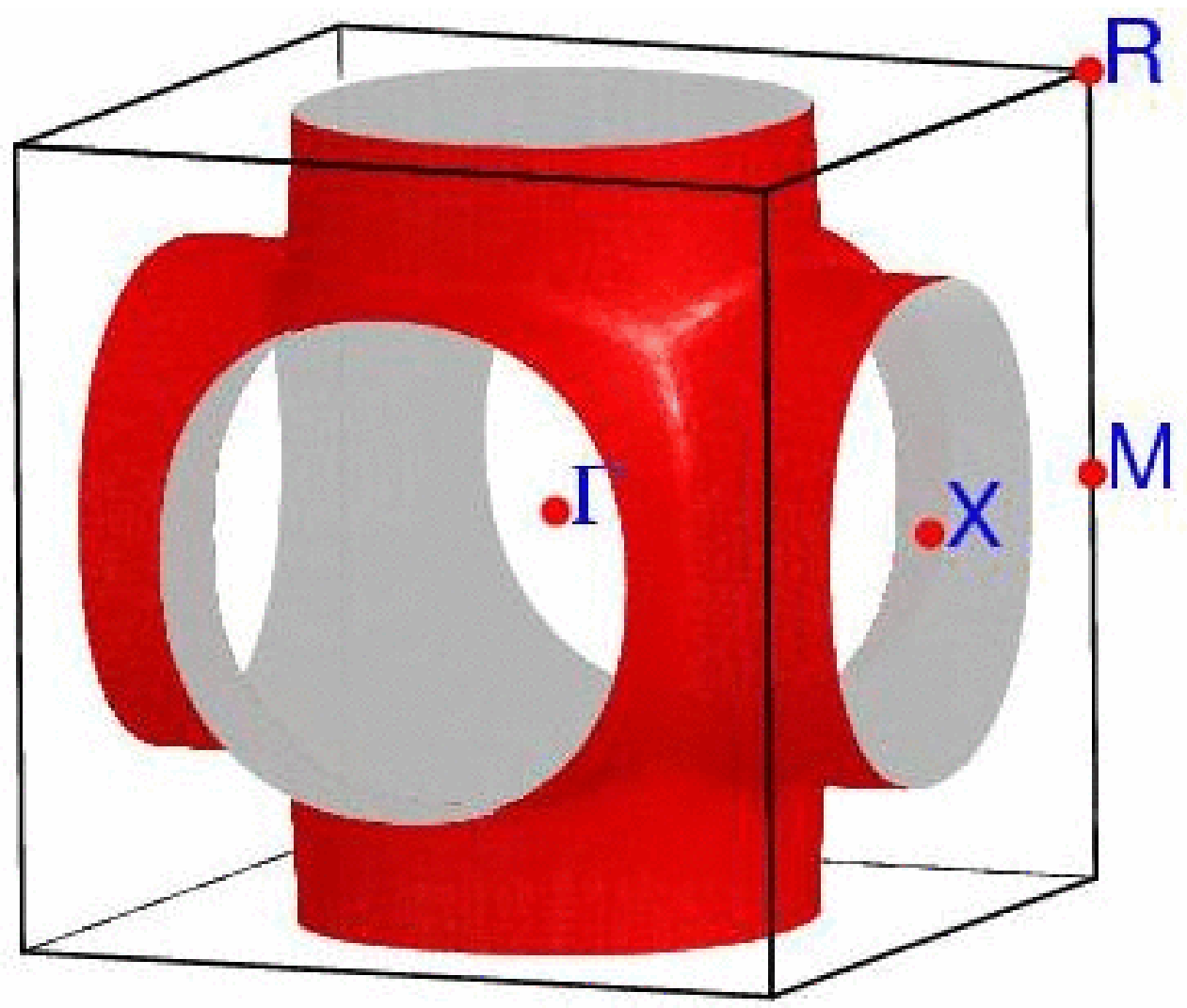}}
\resizebox{6.5cm}{6cm}{\includegraphics{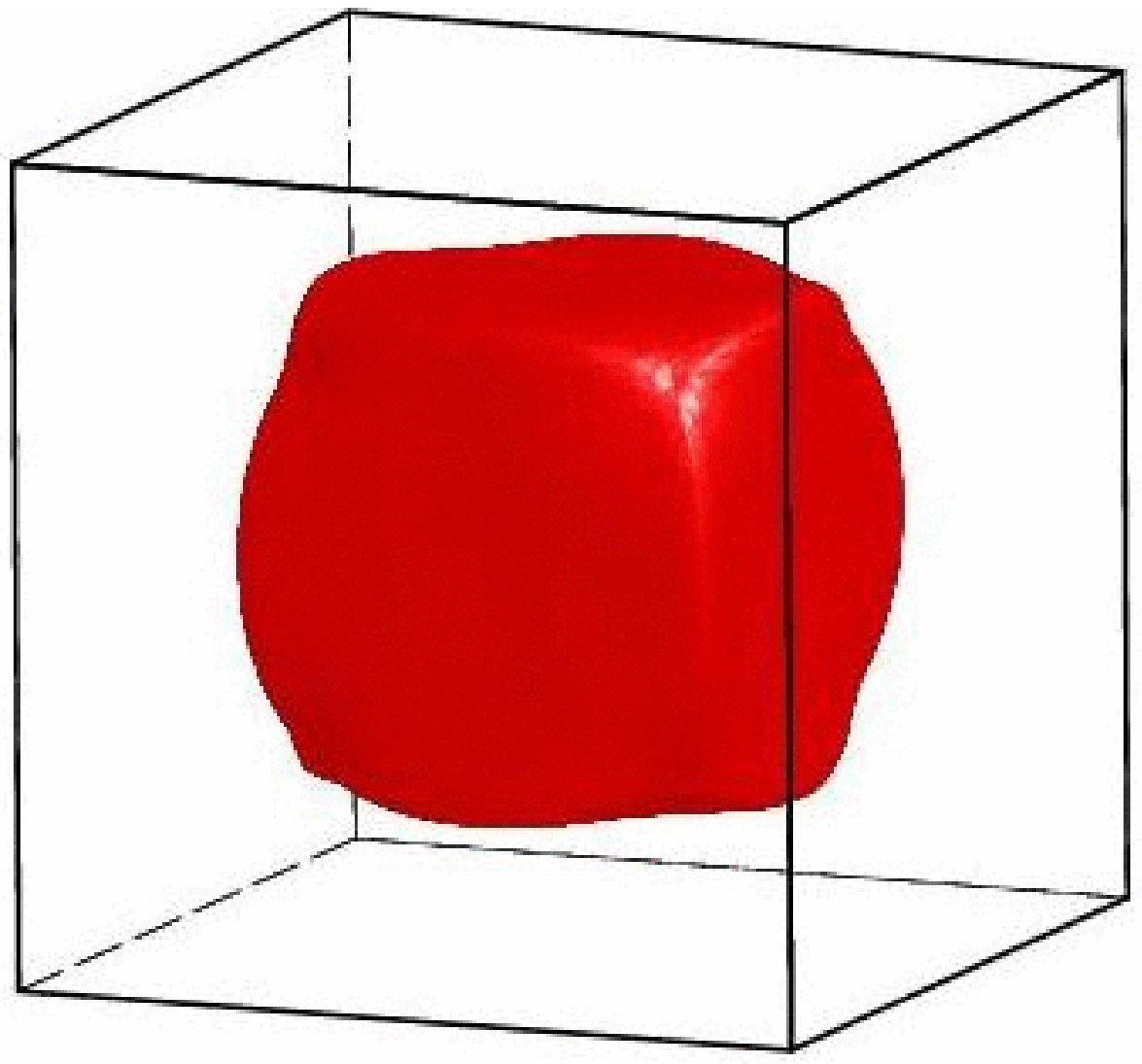}}
\resizebox{6.5cm}{6cm}{\includegraphics{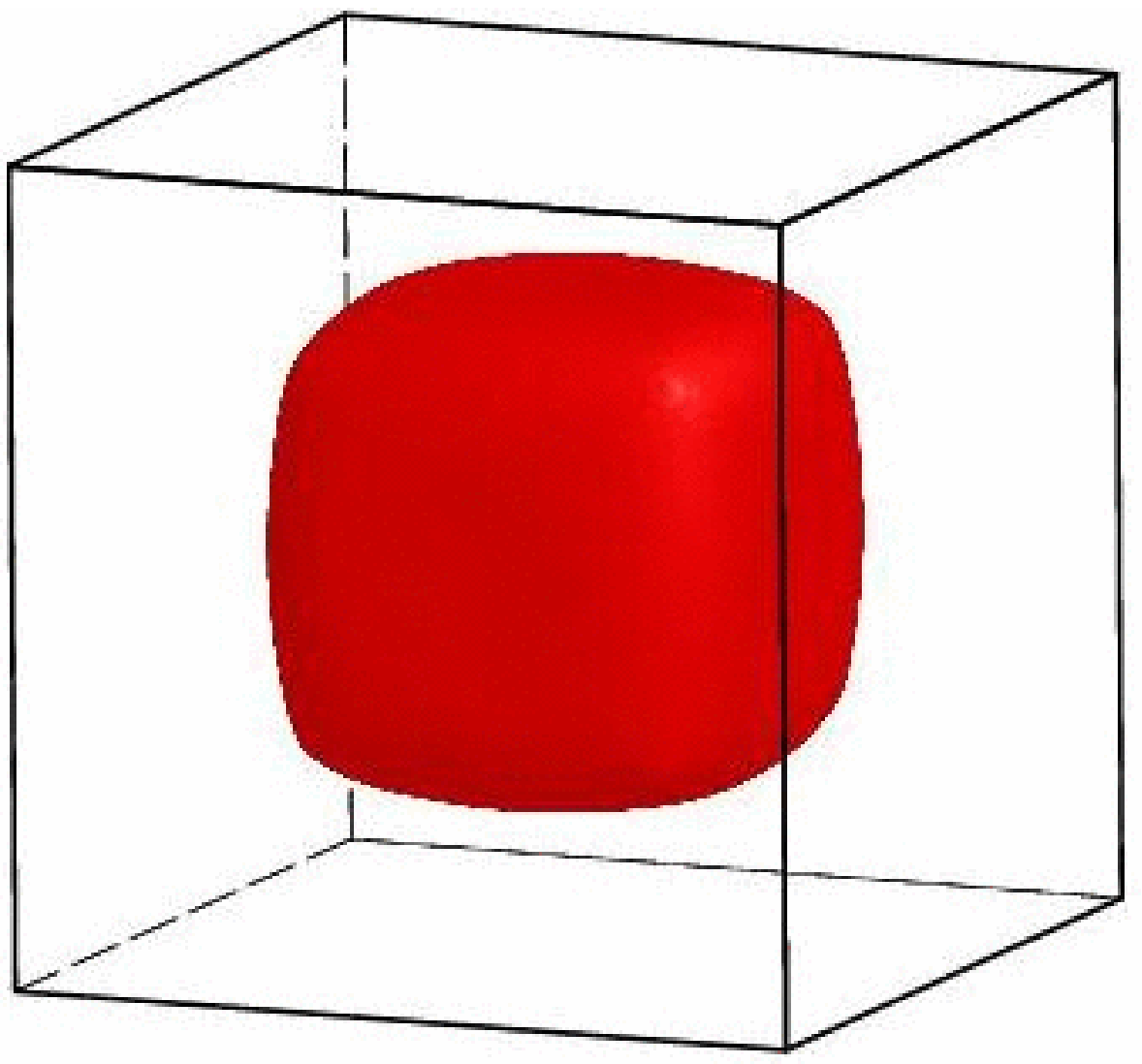}}
\caption{(Color online) Fermi surfaces, which contain electrons
 of NM \sco. 
 These surfaces show nesting features commonly arising in perovskite
 structure transition metal oxides.
 Both the second and the third surfaces are cube-like with
 rounded edges and corners; each face of the second surface is 
 nearly circular.
 The Fermi velocity of 2$\times 10^7$ cm/sec, which is a typical value
 in a metal, is nearly uniform 
 through the surfaces.
}
\label{FS}
\end{figure}

In this subsection we address the electronic structure of 
the NM cubic phase (observed above 40 K) using LDA.
The band structure around the Fermi level E$_F$ (the Cr $d$ regime) 
is shown in Fig. \ref{band}, and the corresponding densities of
states (DOSs) are displayed in Fig. \ref{dos}.
The one-third filled Cr $t_{2g}$ manifold with width of 2 eV 
lies between --1.5 eV and 0.5 eV (we take E$_F$ as the zero of energy).
The unfilled Cr $e_g$ manifold touches the $t_{2g}$ manifold
at the $X$ point at 0.5 eV and extends to 4.5 eV, leading to 
the $t_{2g}$-$e_g$ (midpoint) splitting of roughly 2.5 eV.
Flat bands along the $\Gamma-X-M$ lines result in
a sharp peak in the DOS at 0.5 eV, which otherwise does not have
any distinguishing structures near E$_F$.

The Fermi surfaces shown in Fig. \ref{FS} display nesting
features, indicative of large susceptibilities at related wavevectors and
suggesting the possibility of either magnetic or charge 
instabilities. 
The intersecting pipe-like surface has six circular faces
with a radius of 0.32($\frac{\pi}{a}$), 
so contains $\sim$0.6 electrons per spin. 
Two cube-like surfaces with sides of length $\sim$0.61($\frac{2\pi}{a}$)
are very similar in size, and touch along the $\Gamma-X$ line.

\begin{figure}[tbp]
\resizebox{8cm}{6cm}{\includegraphics{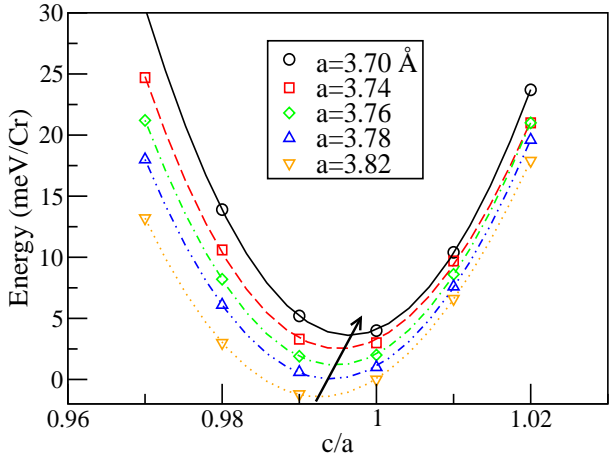}}
\caption{(Color online) Changes in energy SrCrO$_3$ with respect to
lattice parameter ratio ($c/a$) in C-AFM phase, with various choices
of lattice parameter $a$. This $a$ parameter
in the range of 3.70--3.82 \AA~ is approximately the range
between our optimized value in the cubic structure
and the experimentally observed value.
The arrow denotes the change in position of the minima,
pointing out the stronger structural distortion for larger volume.
}
\label{relax}
\end{figure}

\subsection{Structure Relaxation}
Our L(S)DA calculations show this tetragonally distorted structure
is energetically favored over the cubic structure.
As expected from the small change in structure, however, the
difference in energy is small, no more than 1 meV per Cr.
The fact that these two structures are nearly degenerate is consistent
with experimental observations.
The quadrupled supercell, on the other hand, 
has a slightly higher energy than the cubic phase.
In this section, we will focus on only the competing cubic and 
tetragonal phases.

To investigate sensitivity of this structure distortion to volume,
we calculated the energy vs. $c/a$ relation in the range of $a$=3.70--3.82 \AA.
The volume is kept fixed while the $c/a$ ratio of lattice
parameters is varied, since the experiment shows negligible change
in volume between the two observed phases.\cite{attfield2}
The result is given in Fig. \ref{relax}.
In very close agreement with experimental observations, a minimum occurs at $c/a$=0.99
for $a$=3.82 \AA, which is the experimentally observed lattice parameter.
However, with decreasing volume this distortion is gradually relieved, and
finally the cubic structure is favored energetically somewhat below $a\approx$3.70 \AA,
0.06 \AA~ smaller value than our optimized parameter.

\subsection{Energetics}
As discussed in the Introduction, the recent observations indicate 
the coexistence of NM cubic phase and AFM tetragonally distorted 
phase.\cite{attfield2}
Within L(S)DA, AFM order is more favored energetically than
NM in both crystal structures.  In the tetragonal phase, AFM
order ($M$=1.55$\mu_B$)
is favored energetically over FM ($M$=1.16$\mu_B$) by 150 meV/Cr. 
The ferromagnetic state is favored only 11 meV/Cr over the nonmagnetic state.
With Stoner $I=0.6$ eV (see below), the energy gain in a simple Stoner
picture $IM^2/4$ would lead to much larger value of $\sim$0.2 eV/Cr.
Note that the moments are roughly consistent with $S$=1 ions in the presence of strong 
$p-d$ hybridization that is evident from the projected DOS in Fig. \ref{dos}.
The strong reduction in moments from the ionic value also reflects considerable 
hybridization, so
a fixed-local-moment ({\it i.e.} Heisenberg picture) has little use here.
The behavior of ferromagnetic order will be investigated more 
explicitly below by fixed spin moment
calculations.\cite{fsm}
The energy differences between magnetic phases are nearly independent of 
the small distortion that we study here.

\subsection{Fixed Spin Moment (FSM) Studies}
\begin{figure}[tbp]
\resizebox{8cm}{6cm}{\includegraphics{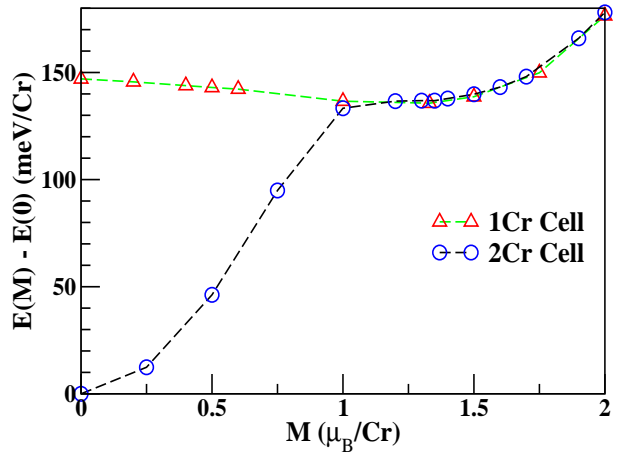}}
\caption{(Color online) Fixed spin moment calculations
 using both one Cr and two Cr cells. 
 The latter cell allows both NM
 and AFM states at moment $M$=0.
 For the doubled cell, the plot shows richer behavior (see text).
Above 1$\mu_B$ the states are the same simple FM aligned state.
}
\label{fsm}
\end{figure}

Initially we used a single Cr cell for our calculations, supplementing
this with doubled cells (see below).
Consistent with our L(S)DA calculations, E(M) has a minimum at $M \sim$1.3 $\mu_B$
and a related small gain of 11 meV in energy, as can be seen in 
curve in Fig. \ref{fsm}.  The very flat E(M) behavior for M up to 1.5$\mu_B$ 
indicates that magnetism in \sco~is very peculiar.  The gain in exchange energy
$I M^2/4$ is almost exactly compensated by a cost in other energy contributions
across this range.
The energy vs. moment curve is fit at small $M$ to the expression
$\varepsilon - \varepsilon_0 = \alpha M^2 + \beta M^4$ to
evaluate the Stoner (exchange) constant $I$ from these FSM calculations.\cite{fsm}
The resulting value of
$\alpha =-17$ meV/$\mu_B^2$ (and $\beta =6$ meV/$\mu_B^4$)
provides the enhanced (observed) susceptibility given by
\begin{eqnarray}
 \chi=\frac{\chi_0}{1-N(0)I}\equiv {\cal S}\chi_0,
\label{chi}
\end{eqnarray}
where the bare susceptibility is $\chi_0=2\mu_B^2N(0)$.
The Stoner enhancement factor ${\cal S}=[2\alpha\chi_0]^{-1}$
is about $-8$. Thus the Stoner $I=0.6$ eV, and $IN(0)=1.1$
with $N(0)=1.87$ states per eV per spin from our FM calculations, 
predicting the system is beyond the Stoner magnetic instability in
the cubic phase (within L(S)DA).

To generalize the study, we used the two-Cr supercell and started
from C-AFM order (G-AFM order showed similar change)
which also has total moment $M$=0. The magnetic field that is applied
in the fixed spin moment method provides an evolution of the C-AFM state
into a FM state, perhaps through an intermediate ferrimagnetic phase.
A small systematic energy difference 
between the single and doubled cell energies that we compare 
has been accommodated by aligning
the M=0 energies. 
We find the  energy vs. $M$ curve to be comprised of three
separate regimes: AFM at $M=0$ where the energy is $\sim$150 meV/Cr lower than
for NM; ferrimagnetic
for $0<M\leq 1$ where the two Cr moments differ; then FM for $M>1$ where both 
calculations describe
the same simple FM phase.
While one might expect strong magneto-elastic coupling in this system,
we find that our FSM results are insensitive to $c/a$ ratio
in the range given in Fig. \ref{relax}.

\begin{figure}[tbp]
\resizebox{8cm}{6cm}{\includegraphics{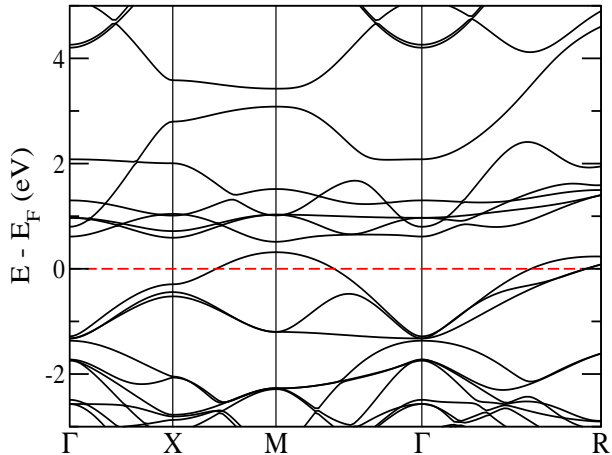}}
\caption{(Color online) Enlarged band structure of C-AFM
in the tetragonally distorted structure, plotted in the basal plane.
That of the cubic phase with the same volume is similar, 
so is not shown.
The antiferromagnetism introduces a gap at the $X$ point in the range of
--0.5-0.5 eV, and the $t_{2g}$ bands become disconnected from the higher
$e_g$ bands due to the reduced bandwidths. 
In unit of $\pi/a$, the $X$ point is (1/2,1/2,0).
}
\label{afmband}
\end{figure}

\begin{figure}[tbp]
\vskip 8mm
\resizebox{8cm}{6cm}{\includegraphics{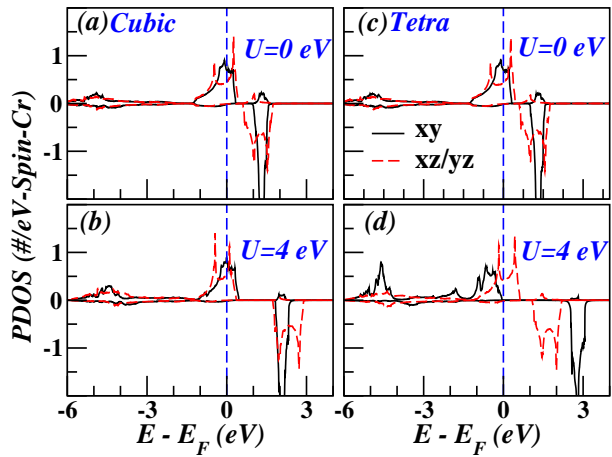}}
\caption{(Color online) $U$-dependent orbital-projected densities
 of states of Cr $t_{2g}$ states for (a)-(b) the cubic and
 (c)-(d) the tetragonal phase, at $U$=0 and 4 eV in C-AFM order.
 At $U$=4 eV, an orbit-ordering transition in $d_{xy}$ orbital
 occurs in the tetragonal phase.
}
\label{udos}
\end{figure}

\subsection{L(S)DA Electronic Structure of the AFM Tetragonally Distorted Phase}
Although Attfield and coworkers observed coexistence of 
the nonmagnetic cubic phase and the antiferromagnetic tetragonal phase,
our L(S)DA calculations show energetically favored AFM in both phases,
as already addressed.
The C-AFM band structure is shown in Fig. \ref{afmband}. 
The Cr local moment is 1.55 $\mu_B$, with negligible dependence 
on this small tetragonal distortion.
The largest effect of the structure distortion on the band structure occurs 
in the maxima at the $M$ and $R$ points, 
downshifting in energy at most 25 meV.
The topmost band crossing $E_F$ has the $d_{xy}$ character.

\section{Inclusion of Correlation Effects}
L(S)DA predicts the C-AFM phase to be considerably lower in energy than the
NM phase, in disagreement with experimental data that suggests the two
phases are nearly degenerate.  
Also L(S)DA cannot produce the partially orbital order experimentally suggested.
We address this discrepancy by including correlation within the LDA+U approach.
On-site Coulomb repulsion $U$ was applied on the Cr ions with AFM order.
$U$ was varied in the range of 0--8 eV, but
the Hund's exchange integral $J$=1 eV is fixed since the results 
in the physical range of $U$ are expected to be insensitive to $J$.

At $U$=0=$J$ (i.e., LSDA level), two electrons are evenly distributed in 
the three bands of the majority $t_{2g}$ manifold.
In the C-AFM (tetragonal) phase, increasing $U$ changes occupancies,
with $d_{xz}$ and $d_{yz}$ weight transferring into $d_{xy}$. 
At $U_c$=4 eV, the majority $d_{xy}$ band 
is fully occupied, while the majority $d_{xz}$ and $d_{yz}$ bands
share equally the other electron,
as shown in the orbital-projected DOS given in Fig. \ref{udos}. 
This $d_{xy}^1$$(d_{xz}d_{yz})^1$ state is consistent with
Attfield and coworkers' conclusion from their neutron diffraction
measurements.\cite{attfield2}
This partial orbitally ordered arrangement remains for larger values of $U$. 
A potentially Mott-insulating $d_{xz}^1, d_{yz}^1, d_{xy}^0$ state is
available but did not arise in the calculations.
In the range of $U$ studied here, this system remains metallic.

An orbital ordering transition 
in a multiband system can result in significantly different
bandwidths.\cite{arita} 
However, in this case, the difference in the occupied bandwidths is small, 
about 100 meV.  The important change is that the center of the $d_{xy}$ band lies 
roughly 1 eV lower than those of the partially occupied $d_{xz}, d_{yz}$ bands. 
It is this difference in the band center ({\it i.e.} the on-site energy)
that drives this transition, 
as happened in Na$_x$CoO$_2$ \cite{nacoo} or V$_2$O$_3$.\cite{v2o3}
Keeping the structure (and symmetry) cubic inhibits such an 
orbital-ordering transition, implying close interplay between
structural distortion (even though tiny) and orbital ordering in LDA+U 
calculations in this system as suggested in CaCrO$_3$.\cite{khomskii2}

\begin{figure}[tbp]
\rotatebox{-90}{\resizebox{6.5cm}{7.5cm}{\includegraphics{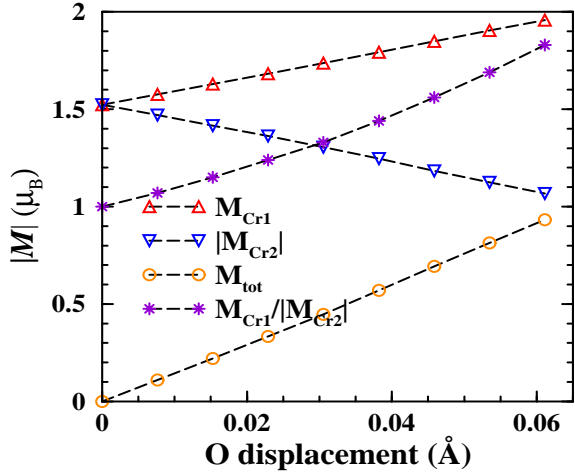}}}
\caption{(Color online) Change in moments due to the O breathing
 vibration when AFM allows.
 The Cr1, Cr2, and total moments are changed 7.1, $-7.5$,
 and 15.3 $\mu_B$/\AA~ in magnitude.
 Note that no displacement leads to AFM.
}
\label{moment}
\end{figure}

\section{O Tilting and Breathing Phonon Modes}
To investigate another possibility of structure distortion
in this perovskite, we used G-AFM order in the quadrupled supercell, 
which allows O tilting and breathing phonon modes in frozen phonon calculations
within L(S)DA.
However, our calculations show these distortions are unfavored
energetically, resulting in stable phonon modes.
These data of change in energy vs. O displacement 
are fit well with a simple harmonic function.
First, for the O tilting vibration 
the phonon energy is 38 meV, typical for
metallic oxides.
For the breathing mode the energy is 89 meV, corresponding to an $rms$
displacement of the oxygen ions by 0.05 \AA.
(Allowing magnetic ordering, both frequencies reduce by $\sim$ 5\%.)

Our calculations show strong magneto-phonon coupling for the breathing mode.
When AFM ordering is included, the Cr local moment
of $M_{Cr}$=1.5 $\mu_B$
is modulated by about $\pm$7.3 $\mu_B$/\AA, as shown in Fig. \ref{moment}.
These changes are quite large, even larger than the change 
in Fe moment, 6.8 $\mu_B$/\AA,  
in LaFeAsO when As ions are displaced, which is widely discussed as
unusually strong magneto-phonon coupling.\cite{zhiping}
At the {\it rms} displacement, the difference between Cr moments becomes 
0.73 $\mu_B$, illustrating just how large the modulation of the moment 
by the breathing mode is.
The Cr charge ``disproportionation'' due to O breathing also 
shows a value of $\pm$0.13$e$ ({\it i.e.} a 0.26$e$ charge difference)
at the {\it rms} displacement,
corresponding to a shift in charge of $\pm2.6e$/\AA.

\section{Summary}
We have presented and analyzed the electronic structure, magnetic ordering, and
the impact of strong correlation effects in the perovskite material
\sco, which is reported in a nonmagnetic
cubic phase coexisting with an antiferromagnetic phase concurrent 
with a small distortion in structure below 40 K.
Consistent with these observations, L(S)DA predicts the slightly distorted
tetragonal structure, but overestimates the polarization energy.
With L(S)DA the cubic phase is magnetically
unstable, with a Stoner
product $IN(0) \approx$1.1.  The Fermi surface shows 
nesting features, but they do not correlate with known data.
Although it is possible for certain crystal symmetries for
orbital ordering to occur withouth changing 
the crystal structure,\cite{khomskii2}
we have not pursued such possibilities.

Including correlated effects within the LDA+U approach, the distorted phase
undergoes an orbital ordering transition at a critical interaction
strength $U_c$=4 eV,
leading to $t_{2g}^2 \rightarrow d_{xy}^1$($d_{xz}d_{yz}$)$^1$ orbital
ordering and structural transition to tetragonal structure.
The structural symmetry lowering is crucial; the cubic phase remains 
a simple metal even for higher $U$.
We have also demonstrated that the O breathing modes show strong 
magneto-phonon coupling.

The magnetic  behavior in \sco~remains unclear.
The experimental data show weak magnetic behavior,\cite{attfield2}
and the susceptibility does not follow Curie-Weiss behavior.\cite{good2}
In contrast to these observations,
our various calculations always result in a full moment 
corresponding to $S$=1 configuration (reduced by hybridization)
as expected for a $d^2$ ion.
Without any peak in the DOS, the temperature variation cannot be modeled
with temperature broadening 
as can be done, for example, in TiBe$_2$.\cite{jeong}

\section{Acknowledgments}
We acknowledge important communications with J. P. Attfield concerning his
experimental observations, J.-S. Zhou for clarifying structure of \cco,
and A. Kyker for illuminating discussion of temperature dependent
susceptibility.
This work was supported by DOE under Grant No. DE-FG02-04ER4611, and interaction
within DOE's Computational Materials Science Network is acknowledged.
K.W.L. was partially supported by a Korea University Grant No. K0718021.

\end{document}